\documentclass[journal]{IEEEtran}
\usepackage{cite}
\usepackage{amsmath,amssymb,amsfonts}
\usepackage{algorithm}
\usepackage{algorithmic}
\usepackage{graphicx}
\usepackage{textcomp}
\usepackage{booktabs}
\usepackage{array}
\usepackage{bm}
\usepackage{hyperref}

\def\BibTeX{{\rm B\kern-.05em{\sc i\kern-.025em b}\kern-.08em
    T\kern-.1667em\lower.7ex\hbox{E}\kern-.125emX}}
\begin{document}
\title{PET Image Reconstruction Using\\
Deep Diffusion Image Prior}
\author{Fumio Hashimoto and Kuang Gong
\thanks{This work was supported by NIH grants R01EB034692 and R01AG078250. (Corresponding author: Kuang Gong)}
\thanks{This work has been submitted to the IEEE for possible publication.}
\thanks{F. Hashimoto and K. Gong are with the J. Crayton Pruitt Family Department of Biomedical Engineering, University of Florida, FL, USA (e-mail: fumio.hashimo@ufl.edu, kgong@bme.ufl.edu). }
\thanks{ }}

\maketitle

\begin{abstract}
Diffusion models have shown great promise in medical image denoising and reconstruction, but their application to Positron Emission Tomography (PET) imaging remains limited by tracer-specific contrast variability and high computational demands. In this work, we proposed an anatomical prior–guided PET image reconstruction method based on diffusion models, inspired by the deep diffusion image prior (DDIP) framework. The proposed method alternated between diffusion sampling and model fine-tuning guided by the PET sinogram, enabling the reconstruction of high-quality images from various PET tracers using a score function pretrained on a dataset of another tracer. To improve computational efficiency,  the half-quadratic splitting (HQS) algorithm was adopted to decouple network optimization from iterative PET reconstruction. The proposed method was evaluated using one simulation and two clinical datasets. For the simulation study, a model pretrained on [$^{\text{18}}\text{F}$]FDG data was tested on [$^{\text{18}}\text{F}$]FDG data and amyloid-negative PET data to assess out-of-distribution (OOD) performance. For the clinical-data validation, ten low‑dose [$^{\text{18}}\text{F}$]FDG datasets and one [$^{\text{18}}\text{F}$]Florbetapir dataset were tested on a model pretrained on data from another tracer. Experiment results show that the proposed PET reconstruction method can generalize robustly across tracer distributions and scanner types, providing an efficient and versatile reconstruction framework for low-dose PET imaging.
\end{abstract}

\section{Introduction}
\label{sec:introduction}
\IEEEPARstart{P}{ositron} emission tomography (PET) is a functional imaging modality that enables \textit{in vivo} visualization and quantification of radiopharmaceutical kinetics. Its intrinsic quantitative capabilities have led to wide applications in oncology, cardiology, and neurology across both clinical and research settings. However, PET images are inherently degraded by limited photon counts received and poor spatial resolution, which compromise quantitative accuracy and lesion detectability \cite{Phelps2006}. Although maximum-likelihood expectation-maximization (MLEM) and its ordered-subset variant, ordered-subset expectation–maximization \cite{lange1984reconstruction,hudson1994accelerated}, have long served as the backbone of PET image reconstruction, high noise levels can obscure clinically important structural details, especially under low-dose conditions. To address this, maximum \textit{a posteriori} (MAP)-based PET reconstruction has been developed, leveraging prior information from other imaging modalities or from the PET image itself to reduce noise and improve resolution \cite{Green1990,gindi1993bayesian, bowsher1996bayesian, Qi1998, Nuyts2002, tang2009bayesian,Bai2013}.

Advances in deep learning have significantly enhanced PET image reconstruction when compared to conventional state‑of‑the‑art algorithms \cite{Kim2018, yang2018artificial, Xie2020, Reader2020, hu2023dynamic, Hashimoto2024}. Early convolutional neural network (CNN)-based methods, such as those by Gong et al.\cite{Gong2019} and Mehranian and Reader\cite{Mehranian2020}, integrated CNNs into iterative PET reconstruction frameworks to improve image quality. More recently, diffusion models have emerged as a powerful alternative \cite{Ho2020,Song2021}. Unlike CNN-based approaches that rely on deterministic input–output mappings, diffusion models are distribution-based methods that can explicitly model image priors, helping to mitigate over-smoothing and achieve more robust image restoration. These models have demonstrated superior performance in PET image denoising applications \cite{Gong2024,xie2023,shen2023pet,Cho2024,pan2024full,Yu2025,webber2025personalized,zhou2024unsupervised,Clementine2025,zhang2024realization}.

For PET image reconstruction, Singh et al. \cite{Singh2024} presented the first demonstration of diffusion model-based PET reconstruction using diffusion posterior sampling (DPS) \cite{Chung2023} and decomposed diffusion sampling (DDS) \cite{Chung2024}. Webber et al. \cite{webber2025likelihood} introduced a likelihood‑scheduled score‑based framework to accelerate reconstruction speed for fully 3D PET reconstruction. Bae et al. \cite{Bae2025} proposed an efficient diffusion model-based PET image reconstruction framework through half-quadratic splitting (HQS) within the RED-Diff framework \cite{mardani2023variational}. While diffusion models show substantial promise for medical image reconstruction, they can experience performance degradation in out-of-distribution (OOD) scenarios, such as data from different scanners or protocols, due to shifts in image characteristics or contrast. In PET imaging, these challenges are amplified by exceptionally high noise levels, pronounced tracer-dependent contrast variability, and the heavy computational demands of fully 3D reconstruction. Consequently, OOD adaptation strategies developed for other imaging modalities \cite{10829716, Chung2024a} have yet to be effectively translated to PET imaging.

 \begin{figure*}[!t]
  \centering
  \includegraphics[width=0.8\linewidth]{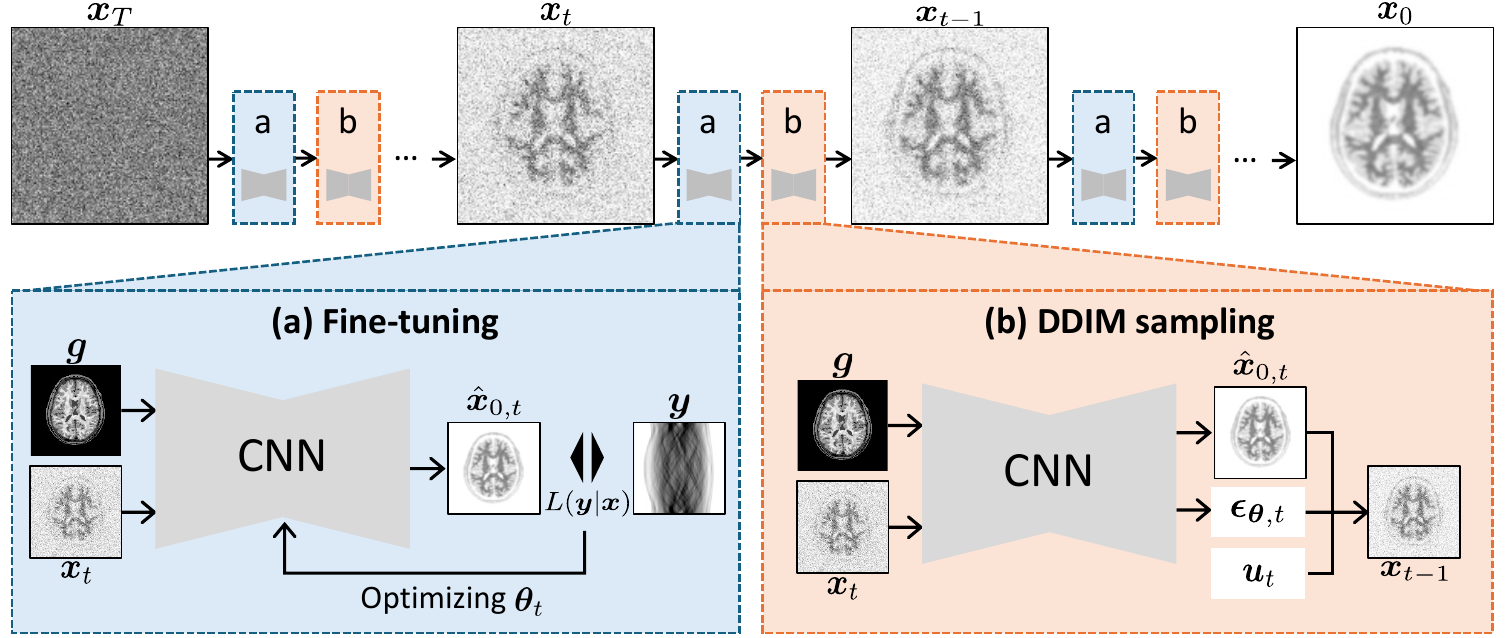}
  \caption{Overview of the proposed DDIP-based and anatomical prior-guided PET image reconstruction framework. At each time step $t$, the proposed method comprises two sub-steps: (a) a fine-tuning sub-step, solved using the HQS; and (b) a DDIM sampling sub-step.}
\label{fig:overview}
\end{figure*}

In this study, we proposed a diffusion model-based, anatomical prior-guided PET image reconstruction method, inspired by the deep image prior (DIP)\cite{Ulyanov2020, Gong2019a, Hashimoto2022} and deep diffusion image prior (DDIP) \cite{Chung2024a} frameworks. The proposed method alternated between diffusion sampling and fine‑tuning a score function using measured PET sinogram data. In addition, the HQS algorithm \cite{Geman1995} was adopted to decouple network optimization and PET reconstruction to enable computationally efficient implementation.

The key contribution of the proposed DDIP-based PET reconstruction method is its ability to reconstruct PET data from various tracers and scanners using a score function pretrained on PET data from a different tracer and/or scanner. Furthermore, we extended the DDIP framework to a conditional model, thereby enabling patient-specific prior guided sampling that could further enhance  PET reconstruction performance. To assess its effectiveness, we first conducted a simulation study in which the score function was trained on simulated [$^{\text{18}}\text{F}$]FDG PET data but tested on an amyloid PET data, a challenging scenario in which tracer distributions differed drastically, i.e., gray–white matter contrast was inverted. We then evaluated the proposed method on clinical datasets, where the score function was pretrained on [$^{\text{18}}\text{F}$]MK‑6240 tau PET scans and tested on [$^{\text{18}}\text{F}$]FDG and [$^{\text{18}}\text{F}$]Florbetapir data acquired from a different PET scanner. These experiments highlight typical OOD adaptation challenges, demonstrating the robustness and generalization capabilities of the proposed DDIP-based PET reconstruction framework.

\section{Background}
\subsection{PET image reconstruction}
PET image acquisition can be modeled through a discrete linear transformation
\begin{equation}
\label{eq:PET}
\bm{\bar{y}} = \bm{Ax}+\bm{b}.
\end{equation}
Here $\bm{\bar{y}} \in \mathbb{R}^{M}$ is the mean of the measured PET data,  $\bm{x} \in \mathbb{R}^{N}$ indicates unknown spatial distribution of the radioactive tracer, ${M}$ is the sinogram size,  $N$ is the image size, $\bm{A}\in \mathbb{R}^{M \times N}$ is the system matrix, and $\bm{b} \in \mathbb{R}^{M}$ represents background components, i.e., random and scattered coincidence events. Supposing the measured PET data $\bm{{y}} \in \mathbb{R}^{M}$ follows a Poission distribution with mean equal to $\bm{\bar{y}}$, we can obtain the log-likelihood function as follows
\begin{equation}
\label{eq:PLL}
L(\bm{y} | \bm{x}) = \sum_{i} y_{i}\log\!\bigl([\bm{A}\bm{x}]_{i} + b_{i}\bigr)
  - \bigl([\bm{A}\bm{x}]_{i} + b_{i}\bigr).
\end{equation}
The MLEM update for voxel $j$ can be obtained as 
\begin{equation}
\label{eq:MLEM}
\hat x^{n+1}_{j,\text{EM}}
= \frac{\hat{x}^{n}_j}{S_j}
  \sum_{i} A_{ij}\,\frac{y_i}{[\bm{A{\hat{x}}}^n]_i + b_i},\quad S_j
=  \sum_{i} A_{ij}.
\end{equation}

\subsection{Diffusion models}
Diffusion models are generative models that learn to approximate a complex data distribution by progressively transforming samples from a simple prior distribution, typically a Gaussian distribution. Suppose that the clean image $\bm{x}_0$ is sampled from the data distribution $q$. The forward diffusion process is defined as a Gaussian Markov chain as follows
\begin{equation}
\label{eq:diffusion1}
q(\bm{x}_{1:T}|\bm{x}_{0}) = \prod_{t=1}^T q\bigl(\bm{x}_t | \bm{x}_{t-1}\bigr),
\end{equation}
\begin{equation}
\label{eq:diffusion2}
q(\bm{x}_t|\bm{x}_{t-1}) = \mathcal{N} (\bm{x}_t; \sqrt{\alpha_t}\bm{x}_{t-1}, \beta_t \bm{I}),
\end{equation}
where $\alpha_t = 1-\beta_t$ and $\beta_t \in (0,1)$ is a predefined variance schedule. In the reverse diffusion process of a denoising diffusion probabilistic model (DDPM) \cite{Ho2020}, sampling can be conditioned using anatomical information $\bm{g}$ as follows
\begin{equation}
\label{eq:reverse}
p_{\theta}\bigl(\bm{x}_{t-1} | \bm{x}_{t}, \bm{g}\bigr)
= \mathcal{N}\bigl(\bm{x}_{t-1};\,\bm{\mu}_{\bm{\theta}}(\bm{x}_{t}, t, \bm{g}) , \sigma_t^2 \bm{I}),
\end{equation}
\begin{equation}
\label{eq:mean}
\bm{\mu}_{\bm{\theta}}(\bm{x}_{t}, t, \bm{g}) = \frac{1}{\sqrt{\alpha_t}}\left[\bm{x}_t - \frac{\beta_t}{\sqrt{\bar{\beta}_t}}\bm{\epsilon}_{\bm{\theta}}(\bm{x}_{t}, t, \bm{g})\right],
\end{equation}
where $\bar\alpha_t= \prod_{s=1}^t\alpha_s$ and $\bar{\beta}_t = 1 - \bar{\alpha}_t$. $\bm{\mu}_{\bm{\theta}}$ and $\bm{\epsilon}_{\bm{\theta}}$ are the mean of the reverse process and the score function, respectively, modeled by a neural network with parameter $\bm{\theta}$. $\sigma_t^2=\frac{1-\bar\alpha_{t-1}} {1-\bar\alpha_t}\beta_t$ is the posterior variance. Based on the trained score function $\bm{\epsilon}_{\hat{\bm{\theta}}}$, the refinement step becomes
\begin{equation}
\label{eq:reverse_sample}
\bm{x}_{t-1} = \frac{1}{\sqrt{\alpha_t}}\left[\bm{x}_t - \frac{\beta_t}{\sqrt{\bar{\beta}_t}}\bm{\epsilon}_{\hat{\bm{\theta}}}(\bm{x}_{t}, t, \bm{g})\right] + \sigma_t\bm{u}_t, 
\end{equation}
where $\bm{u}_t \sim \mathcal{N}(\bm{0},\bm{I})$. 

Denoising diffusion implicit model (DDIM) \cite{Song2021a} replaces the stochastic reverse chain with a deterministic update using Tweedie’s formula to estimate a clean image $\bm{\hat{x}}_0$ at time $t$,
\begin{align}
\label{eq:tweedie}
\hat{\bm{x}}_0(\bm{x}_t, t, \bm{g}) &=\bm{x}_t + \sigma_t^2 \nabla_{\bm{x}_t} \log p_{\hat{\bm{\theta}}}\bigl(\bm{x}_t | \bm{g}\bigr)\notag\\
&\approx \bm{x}_t + \sigma_t^2\bm{\epsilon}_{\hat{\bm{\theta}}}(\bm{x}_t, t, \bm{g}).
\end{align}
The refinement step during reference diffusion becomes
\begin{align}
\label{eq:ddim}
\bm{x}_{t-1}
= &\sqrt{\bar{\alpha}_{t-1}}\hat{\bm{x}}_0(\bm{x}_t,t, \bm{g})\notag\\
&+ \sqrt{\bar{\beta}_{t-1} - \eta^2\bar{\beta}_t} \bm{\epsilon}_{\hat{\bm{\theta}}}(\bm{x}_t, t, \bm{g})\notag\\
&  + \eta\sigma_t\bm{u}_t,
\end{align}
where  $\eta \in [0,1]$ is a hyperparameter controlling the stochasticity. Note that DDIM sampling is the same as the stochastic DDPM sampling when $\eta = 1.0$.

\begin{table}[!t]
\centering
\caption{Combinations of components for DDIP used in the ablation study.}
\label{table_ablation}
\setlength{\tabcolsep}{3pt}
\begin{tabular}{lccc}
\toprule
Methods& 
Diffusion& 
Anatomical information&LoRA\\\midrule
(a) cDIP& 
& 
$\checkmark$&\\
(b) uDDIP& 
$\checkmark$& 
&$\checkmark$\\
(c) cDDIP w/o LoRA& 
$\checkmark$&
$\checkmark$&\\
(d) cDDIP (Proposed)& 
$\checkmark$& 
$\checkmark$&$\checkmark$\\ 
\bottomrule 
\end{tabular}
\end{table}

 \begin{figure}[!t]
  \centering
  \includegraphics[width=1.0\linewidth]{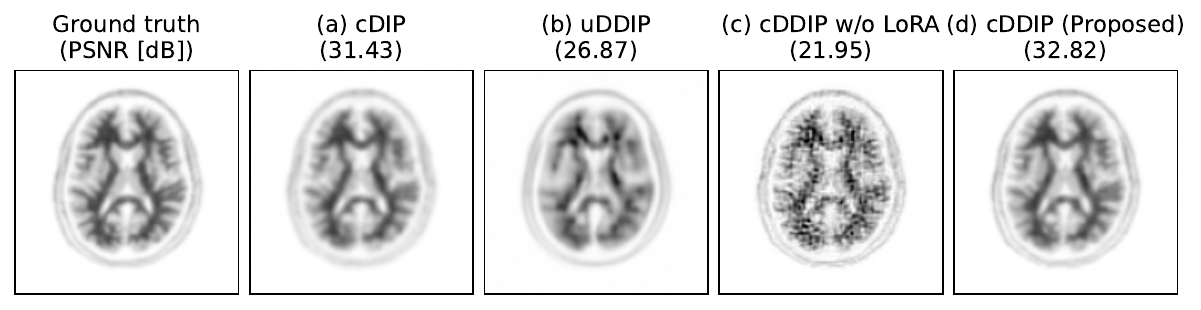}
  \caption{Ablation study on the combinations of components for DDIP.}
\label{fig:ablation}
\end{figure}

 \begin{figure}[!t]
  \centering
  \includegraphics[width=1.0\linewidth]{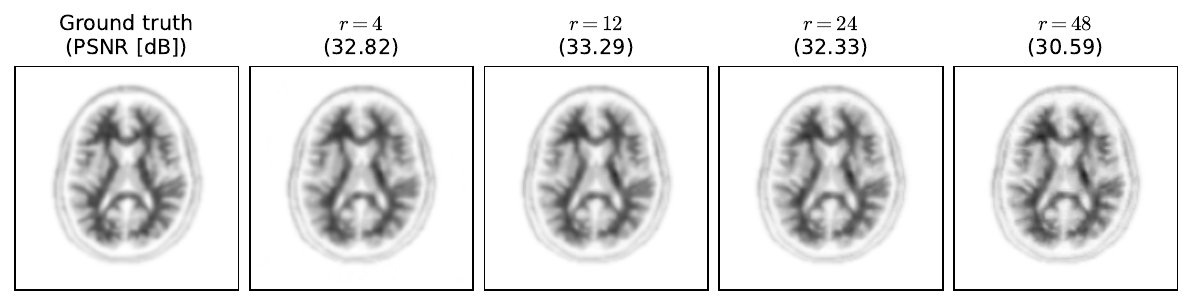}
  \caption{Simulation results of amyloid-negative data under different LoRA configurations, including varying rank values $r$.}
\label{fig:lora-img}
\end{figure}

 \begin{figure*}[!t]
  \centering
  \includegraphics[width=\textwidth]{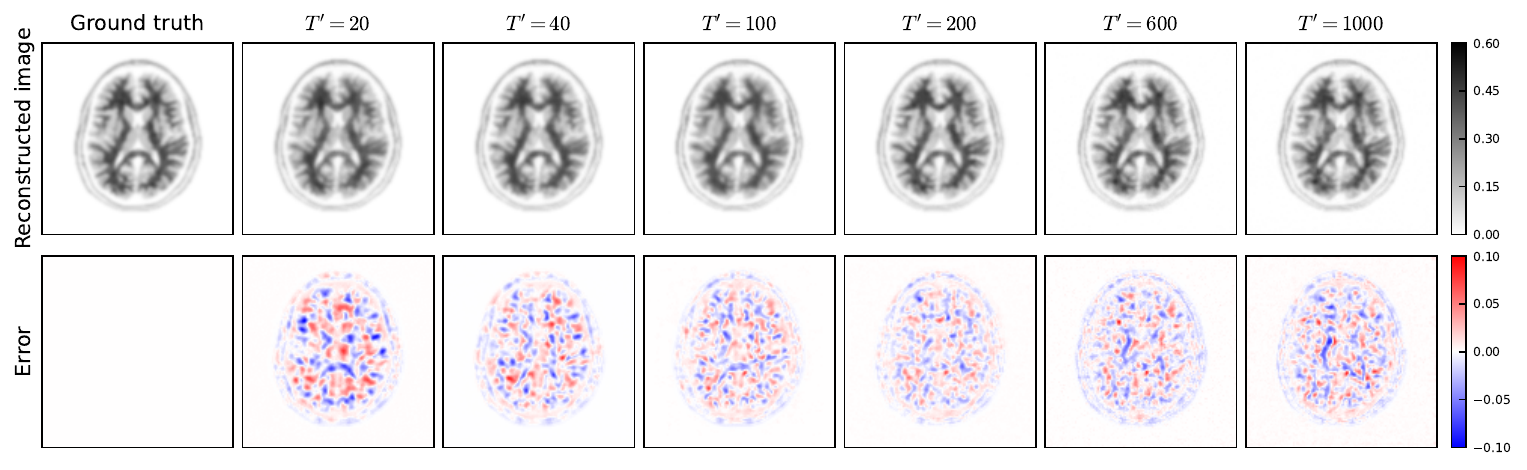}
  \caption{Simulation results of the amyloid-negative data for different starting times $T'$. Rows correspond to the reconstructed images (top) and the error maps (bottom), where each error map is defined as target image - ground truth. The grayscale bar at the top indicates activity (a.u.); and the bottom color bar indicates signed differences (a.u.).}
\label{fig:simu_beta_img}
\end{figure*}

 \begin{figure}[!t]
  \centering
  \includegraphics[width=0.85\linewidth]{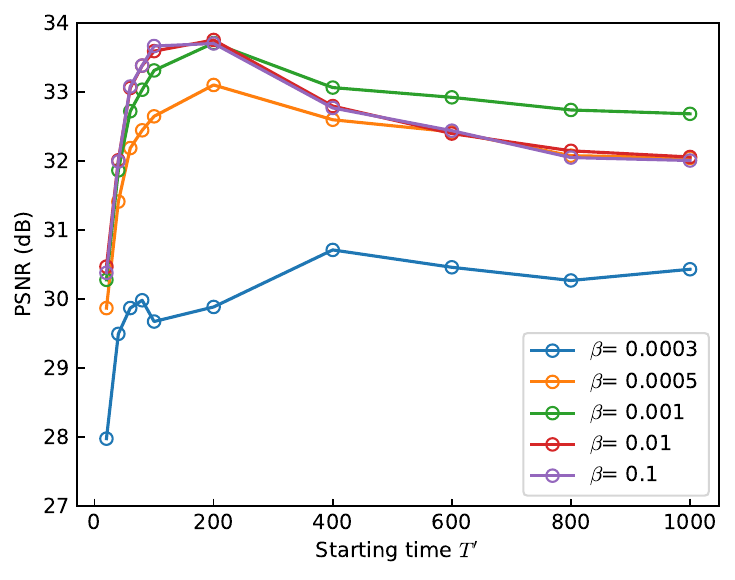}
  \caption{Effect of the hyperparameter $\beta$ on PSNR for the simulation data.}
\label{fig:beta}
\end{figure}

\subsection{DDIP for OOD adaptation}
In diffusion model-based inverse problems such as medical image reconstruction, the score function is trained on images acquired from a specific contrast and scanner configuration. As a result, performance may degrade in OOD scenarios. Steerable conditional diffusion \cite{10829716} and its generalization DDIP \cite{Chung2024a} addressed the limitations by fine-tuning a pre-trained score function using the measured data, inspired by the DIP framework \cite{Ulyanov2020}, as follows
\begin{equation}
\label{eq:ddip}
\hat{\bm{\theta}} = \underset{\bm{\theta}}{\arg\min} \lVert\bm{y} - \bm{A}\bm{\hat{x}}_0(\bm{x}_t, t)\lVert_2^2 \quad \text{for } t = T, T-1, \ldots, 1.
\end{equation}
After fine‐tuning at time $t$, the updated score function generates the sample $\bm{x}_{t-1}$ using the DDIM update in (\ref{eq:ddim}). Note that although we have so far described diffusion models in a conditional setting, the original DDIP framework proposed is purely unconditional \cite{Chung2024a} . When $t=T$, Equation~(\ref{eq:ddip}) reduces to the original DIP optimization.

\section{Methodology}
\subsection{Proposed framework}
An overview of the proposed DDIP-based, anatomical prior-guided PET image reconstruction is shown in Fig \ref{fig:overview}. For the proposed method, apart from PET sinogram, the unknown PET image $\bm{x}$ is estimated by also leveraging the patient's own anatomical prior $\bm{g}$, such as magnetic resonance (MR) images. For each time step $t$ during the reverse process, two sub-steps are involved: a fine-tuning sub-step and a DDIM sampling sub-step. At each time step $t$, we first perform the fine-tuning sub-step by minimizing the following objective function
\begin{equation}
\label{eq:ddiprecon}
\hat{\bm{\theta}} = \underset{\bm{\theta}}{\arg\min}\; -L\left( \bm{y} | \hat{\bm{x}}_0(\bm{x}_t, t, \bm{g}) \right). 
\end{equation}
Here $L(\cdot|\cdot)$ is the Poisson log-likelihood function defined in  (\ref{eq:PLL}). Then, in the DDIM sampling sub-step,  $\bm{x}_{t-1}$ is generated based on~(\ref{eq:ddim}). The final reconstructed image $\bm{\hat{x}}_0$ is obtained by repeating the above two sub-steps at each time step down to $t=1$. The fine‐tuning step in (\ref{eq:ddiprecon}) can be interpreted as the conditional DIP (cDIP) optimization \cite{Cui2019PETLearning} at each noise scale along the reverse diffusion process. Therefore, the final estimate $\hat{\bm{x}}_0$ aligns with the measured PET data $\bm{y}$.

As the system matrix and neural network are coupled in~(\ref{eq:ddiprecon}), we decouple PET reconstruction and network optimization using the HQS algorithm \cite{Geman1995}. The fine-tuning sub-step as shown in~(\ref{eq:ddiprecon}) can thus be solved by alternating minimization as
\begin{equation}
\label{eq:hqs1}
\bm{x}^{(n+1)} 
= \underset{x}{\arg\min}-L\bigl(\bm{y} | \bm{x}\bigr)
  +\frac{\beta}{2}\Bigl\lVert \bm{x} - \hat{\bm{x}}_0(\bm{x}_t, t, \bm{g})\Bigr\rVert^2,
\end{equation}
\begin{equation}
\label{eq:hqs2}
\bm{\theta}^{(n+1)} 
= \underset{\bm{\theta}}{\arg\min}\;\Bigl\lVert \bm{x}^{(n+1)} 
  -\hat{\bm{x}}_0(\bm{x}_t, t, \bm{g})\Bigr\rVert^2.
\end{equation}

 \begin{figure*}[!t]
  \centering
  \includegraphics[width=\textwidth]{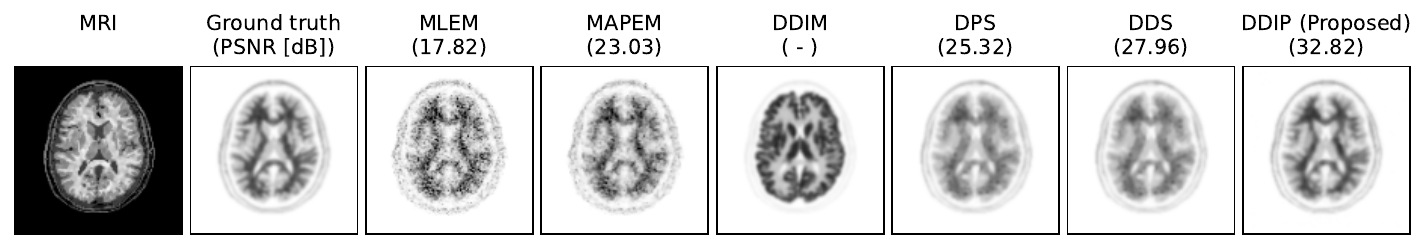}
  \caption{Simulation results of the amyloid-negative data using a model pretrained on the simulated [$^{\text{18}}\text{F}$]FDG dataset.}
\label{fig:simu_img}
\end{figure*}

 \begin{figure*}[!t]
  \centering
  \includegraphics[width=0.9\textwidth]{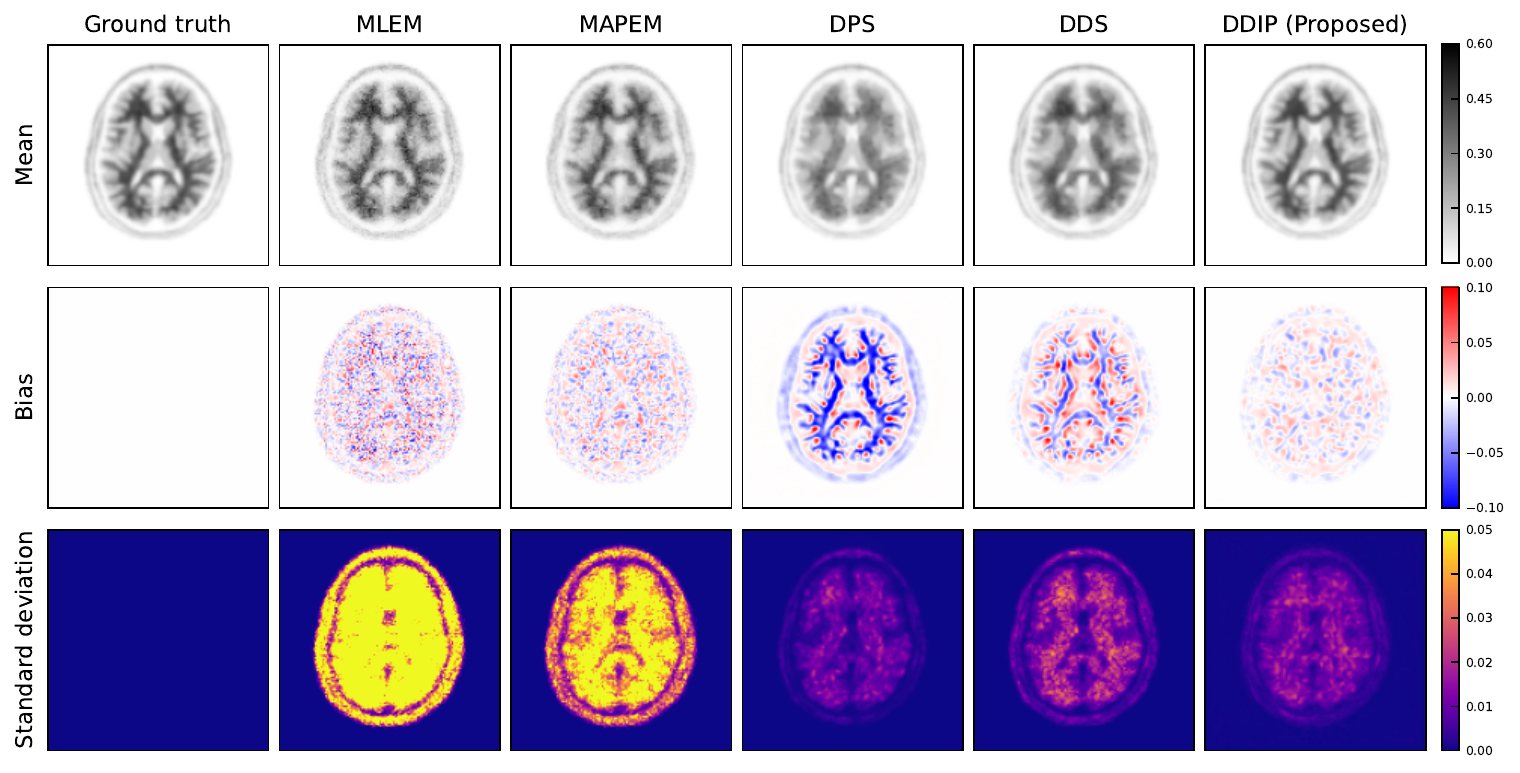}
  \caption{Mean, bias, and standard deviation images for the amyloid-negative simulation data, computed over 10 independent realizations. Rows correspond to: mean images (top), bias maps (middle; defined as target image - ground truth), and standard deviation maps (bottom). The grayscale bar at the top indicates activity (a.u.); the middle color bar indicates signed differences (a.u.); and the bottom color bar indicates the standard deviation (a.u.).}
\label{fig:simu_mean_bias_std_img}
\end{figure*}

\subsection{Solving the image update in (\ref{eq:hqs1}) }
The optimization transfer method \cite{Langed2000, Wang2012} is used to solve the image update in~(\ref{eq:hqs1}). We construct a surrogate function $\mathcal{Q}_L\bigl(\bm{x} | \bm{x}^{(n)}\bigr)$ for $L(\bm{y} | \bm{x})$ to optimize each voxel independently, as follows:
\begin{equation}
\label{eq:surrogate}
\mathcal{Q}_L\bigl(\bm{x} | \bm{x}^{(n)}\bigr)
= \sum_{j} S_j\Bigl(\hat{x}_{j,\text{EM}}^{(n+1)}\log x_j - x_j\Bigr),
\end{equation}
where $\hat{\bm{x}}_{\mathrm{EM}}^{(n+1)}$ and $\bm{S}$ are the MLEM update and the sensitivity image in~(\ref{eq:MLEM}), respectively. The surrogate function can be shown to satisfy the following two properties:
\begin{equation}
\label{eq:inequality1}
\mathcal{Q}_L\bigl(\bm{x} | \bm{x}^{(n)}\bigr)-\mathcal{Q}_L\bigl(\bm{x}^{(n)} | \bm{x}^{(n)}\bigr)
\le L\bigl(\bm{y} | \bm{x}\bigr) - L\bigl(\bm{y} | \bm{x}^{(n)}\bigr),
\end{equation}
\begin{equation}
\label{eq:inequality2}
\nabla \mathcal{Q}_{L}\bigl(\bm{x}^{(n)} | \bm{x}^{(n)}\bigr)
= \nabla L\bigl(\bm{y} | \bm{x}^{(n)}\bigr).
\end{equation}
By combining~(\ref{eq:surrogate}) with the quadratic penalty term in~(\ref{eq:hqs1}), we obtain the final surrogate objective function as
\begin{align}
\label{eq:surrogateobjective}
P\bigl(x_j | \bm{x}^{(n)}\bigr)
=& S_j\Bigl(\hat{x}_{j,\text{EM}}^{(n+1)} \log x_j - x_j\Bigr)\notag\\
&- \frac{\beta}{2}\Bigl[x_j - \hat{x}_0(\bm{x}_t, t, \bm{g})_j\Bigr]^2.
\end{align}
Finally, the voxel-wise closed-form solution for~(\ref{eq:hqs1}) can be obtained by solving ${\partial P} / {\partial x_j} = 0$ in~(\ref{eq:surrogateobjective}) as
\begin{align}
\label{eq:closed}
x_j^{(n+1)}
=& \frac{1}{2}\Biggl[\hat{x}_0(\bm{x}_t, t, \bm{g})_j - \frac{S_j}{\beta}\Biggr]\notag\\
&+ \frac{1}{2}\sqrt{
  \Bigl(\hat{x}_0(\bm{x}_t, t, \bm{g})_j - \frac{S_j}{\beta}\Bigr)^2 + 4\hat{x}_{j,\text{EM}}^{(n+1)}\frac{S_j}{\beta}
}.
\end{align}

\begin{algorithm}[!t]
  \caption{The proposed DDIP-based PET reconstruction}
  \label{alg:ddiprecon}
  \begin{algorithmic}[1]
    \REQUIRE Outer iteration number $N$, inner iteration numbers $M_1$ and $M_2$, start time step $T'$, network parameter $\bm{\theta}$, measured data $\bm{y}$, prior $\bm{g}$
    \STATE $ \hat{\bm{x}}_{0,\bm{\theta}}\bigl(\bm{x}_t,t,\bm{g}\bigr)
      := \bm{x}_t + \sigma_t^2\bm{\epsilon}_{\bm{\theta}}(\bm{x}_t, t, \bm{g}) $
    \STATE Running MLEM with 20 iterations to obtain $\bm{\hat{x}}_{\text{EM-init}}$
    \STATE $\bm{x}_{T'} = \sqrt{\bar{\alpha}_{T'}}\hat{\bm{x}}_{\text{EM-init}} + \sqrt{\bar{\beta}_{T'}}\bm{u}_{T'}$
    \STATE $\bm{\theta}_{T'} =\bm{\theta}$
    \FOR{$t = T'$ {\bfseries to} $1$}
        \FOR{$n= 1$ {\bfseries to} $N$}
        \STATE $\bm{x}^{(n,0)} = \bigl[\hat{\bm{x}}_{0,\bm{\theta}_t}(\bm{x}_t, t, \bm{g})\bigr]_+$
            \FOR{$m= 1$ {\bfseries to} $M_1$}
            \STATE $\hat x^{(n,m)}_{j,\text{EM}} = \frac{x^{(n,m-1)}_j}{S_j} \sum_{i} A_{ij}\frac{y_i}{[\bm{Ax}^{(n,m-1)}]_i + b_i}$
            \STATE $\begin{aligned}
            x_j&^{(n,m)}=\frac{1}{2}\Biggl[x^{(n,0)}_j - \frac{S_j}{\beta}\Biggr]\\
&+ \frac{1}{2}\sqrt{
  \Bigl(x^{(n,0)}_j - \frac{S_j}{\beta}\Bigr)^2 + 4\hat{x}_{j,\text{EM}}^{(n,m)}\frac{S_j}{\beta}}
\end{aligned}$
            \ENDFOR
        \STATE Running optimization with $M_2$ iterations to update\\
        $\bm{\theta}_t = \underset{\bm{\theta}}{\arg\min}\Bigl\lVert \bm{x}^{(n,M_1)} -\hat{\bm{x}}_{0}(\bm{x}_t, t, \bm{g})\Bigr\rVert^2$
        \ENDFOR
    \STATE Running DDIM sampling to obtain\\
        $\begin{aligned}
        \bm{x}_{t-1} =& \sqrt{\bar{\alpha}_{t-1}}\hat{\bm{x}}_{0}(\bm{x}_t,t, \bm{g})\\
&+ \sqrt{\bar{\beta}_{t-1} - \eta^2\bar{\beta}_t} \bm{\epsilon}_{\bm{\theta}}(\bm{x}_t, t, \bm{g})  + \eta\sigma_t\bm{u}_t
\end{aligned}$
    \STATE $\bm{\theta}_{\,t-1} = \bm{\theta}_t$
    \ENDFOR
    \STATE \textbf{return} $\hat{\bm{x}}_0 = \hat{\bm{x}}_{0,\bm{\theta}_{1}}(\bm{x}_1, 1, \bm{g})$
  \end{algorithmic}
\end{algorithm}

\subsection{Solving the network update in (\ref{eq:hqs2}) }
The network update in (\ref{eq:hqs2}) can be solved using a typical deep image prior-based denoising optimization framework \cite{Ulyanov2020}. In this study, we employ a 2D U‐Net architecture with attention and residual blocks used for PET image denoising \cite{Gong2024}; the network details are provided in \cite{Dhariwal2021}. A three-channel input, consisting of the target axial slice and its two adjacent neighbors, is used to mitigate artifacts along the axial direction.

Although the network employed in this study is 2D-based, 3D reconstruction is enabled by extracting overlapping three-slice stacks from the 3D volume and feeding them as individual samples along the batch dimension. The processed stacks are then reassembled into the fully 3D volume along the axial dimension.

Additionally, we introduce low-rank adaptation (LoRA) \cite{Hu2022} during network optimization, allowing adaptation with a reduced number of trainable network parameters while preserving the pre-trained model’s knowledge. In LoRA, each weight matrix $\bm{W}_0 \in \mathbb{R}^{d \times k}$ of the pretrained model is frozen, and instead a low-rank update $\Delta \bm{W}$ is added as follows:
\begin{equation}
\label{eq:lora}
\bm{W}(\bm{\theta}) = \bm{W}_0 + \Delta \bm{W}(\bm{\theta}), 
\quad \Delta \bm{W}(\bm{\theta}) = \bm{U}(\bm{\theta})\bm{V}(\bm{\theta}),
\end{equation}
where $\bm{U} \in \mathbb{R}^{d \times r}$ and $\bm{V} \in \mathbb{R}^{r \times k}$ are trainable low-rank factors parameterized by $\bm{\theta}$, with rank $r \ll \min(d, k)$.  $\bm{U}$ and  $\bm{V}$ are initialized randomly and to zero, respectively; thus, the initial weight matrix $\bm{W}$ is $\bm{W}_0$. LoRA drastically reduces the number of trainable parameters from $d \times k$ to $r(d + k)$, enabling efficient optimization with minimal computational overhead. In this study, we set $r=4$ in all experiments except those varying the LoRA parameter; this value corresponds to only 1.13\% of the model’s total parameters. AdamW \cite{loshchilov2018adamw} is used as the optimizer.

\subsection{Overall algorithm}
The overall algorithm flowchart of the proposed DDIP-based PET reconstruction framework is shown in Algorithm \ref{alg:ddiprecon}. Although the pretrained score function is based on $T=1000$, we accelerate the proposed reconstruction method by starting from $t=T'$. The initial image $\bm{x}_{T'}$ is calculated as:
\begin{equation}
\label{eq:initialize}
\bm{x}_{T'} = \sqrt{\bar{\alpha}_{T'}}\hat{\bm{x}}_{\text{EM-init}} + \sqrt{\bar{\beta}_t}\bm{u}_{T'},
\end{equation}
where $\hat{\bm{x}}_{\text{EM-init}}$ is the initial image reconstructed by the MLEM. To keep the image $\hat{\bm{x}}_{\text{EM-init}}$ relatively less noisy, we set the number of MLEM iterations to 20. In the proposed framework shown in Algorithm \ref{alg:ddiprecon}, $N=2$, $M_1=5$, and $M_2=1$ are adopted.

\section{Experimental Setup}
We evaluated the proposed DDIP-based method on 2D simulation data and real 3D clinical data. For the 2D simulation, we used [$^{\text{18}}\text{F}$]FDG data for training and amyloid data for testing to evaluate an OOD scenario. For the clinical 3D experiment, we assessed performance on two testing datasets, [$^{\text{18}}\text{F}$]FDG and [$^{\text{18}}\text{F}$]Florbetapir, using a score function pretrained on [$^{\text{18}}\text{F}$]MK-6240 tau scans acquired with a different scanner. The simulation study was run on an NVIDIA A100 GPU with 80 GB memory, and the clinical studies were run on an NVIDIA B200 GPU with 192 GB  memory. 
\subsection{Brain phantom simulation study}
We employed twenty BrainWeb phantom datasets \cite{Aubert-Broche2006} in the simulation study. For each phantom, the corresponding T1-weighted MR image was used as the anatomical prior. We divided the 20 subjects into 18 for training, 1 for validation, and the remaining 1 for testing. To evaluate the proposed method under both in-distribution (ID) and OOD scenarios, [$^{\text{18}}\text{F}$]FDG phantoms were generated for training and validation, and one [$^{\text{18}}\text{F}$]FDG phantom and one amyloid-negative phantom with reversed gray–white matter contrast were used for testing. The contrast of the gray matter and white matter was set to $1.0:0.25$ for the [$^{\text{18}}\text{F}$]FDG phantoms and $1.0:3.3$ for the amyloid-negative phantom. The Siemens Biograph mMR scanner was selected as the basis for the modeled geometry. The image size was $128 \times 128$ with a voxel size of $2.08 \times 2.08 \text{ mm}^2$. The system matrix was implemented using ParallelProj \cite{Schramm2023}.

We pretrained the score function by minimizing the following loss function:
\begin{equation}
\label{eq:loss}
L({\bm{\theta}}) = \mathbb{E}_{t\sim{[1,T]},\bm{x}_0\sim q(\bm{x}_0),\bm{\epsilon}\sim\mathcal{N}(\bm{0},\bm{I})} \bigl[\lVert \bm{\epsilon} - \bm{\epsilon}_{\theta}(\bm{x}_t,t) \rVert_2^2 \bigr],
\end{equation}
where $\mathbb{E}$ denotes expectation, $T$ was set to 1000, and $t$ was sampled uniformly. This learns the score of the training data distribution. The proposed method then enforces data consistency, while retaining the same noise schedule and score parameterization as in pretraining, to adapt the pretrained prior to the PET data. AdamW is used as the optimizer.

During training, data augmentation was applied to achieve a 200-fold expansion. This included randomly scaling gray- and white-matter uptake values by factors sampled from $[0.8, 1.2]$, along with applying affine transformations: uniform scaling by factors from $[0.9, 1.05]$, random rotations from $[-15^\circ, 15^\circ]$, and shearing from $[-0.15, 0.15]$. Although [$^{\text{18}}\text{F}$]FDG brain exhibits a gray-to-white matter ratio around 2.5 to 5.0 \cite{BERTI2014129, diagnostics13132254}, ratios can exceed 5 in focal hypermetabolism. Thus, we used a conservative upper bound on uptake scaling in the augmentation. 

During testing, 10 independent and identically distributed (i.i.d.) realizations were generated for the same amyloid-negative phantom. Poisson noise was added at a count level of $4.0\times10^7$. Attenuation, scatter, and randoms were not included for simplicity.

Regarding quantification, the peak signal-to-noise ratio (PSNR) was adopted and calculated as 
\begin{equation}
\label{eq:psnr}
\text{PSNR} = 10\log_{10}\Biggl[\frac{\max{(\bm{K})}^2}{\frac{1}{N}\bigl\lVert\bm{K}-\bm{K}'\bigr\rVert_{2}^{2}}\Biggr],
\end{equation}
where $\max{(\cdot)}$ indicates the maximum value of the image, $\bm{K}$ and $\bm{K}'$ denote the ground truth (phantom) and target reconstructed images, respectively, and $N$ is the number of voxels. The \%contrast and coefficient of variation (CV) were also adopted and calculated as  \cite{Ikari2016}
\begin{equation}
\label{eq:contrast}
\text{\%contrast} = \frac{(\text{GM}_t/\text{WM}_t-1)}{(\text{GM}_{p}/\text{WM}_{p}-1)} \times 100,
\end{equation}
\begin{equation}
\label{eq:cv}
\text{CV} = \frac{\text{SD}_{\text{WM}}}{\text{Mean}_{\text{WM}}},
\end{equation}
where $\text{GM}_t$ and $\text{WM}_t$ are the mean gray matter and white matter values in target reconstructed images, and $\text{GM}_{p}$ and $\text{WM}_{p}$ are the corresponding mean values in the ground-truth images.

\begin{figure}[!t]
  \centering
  \includegraphics[width=0.85\linewidth]{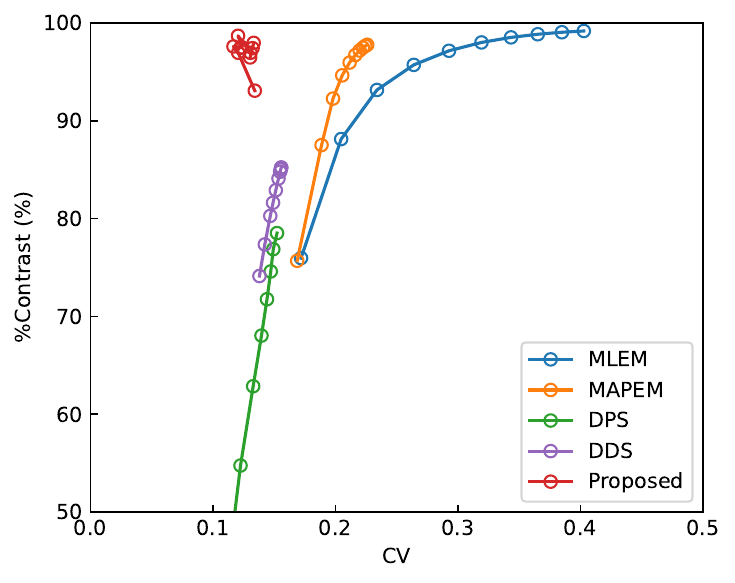}
  \caption{Mean \%contrast-CV tradeoff curves of the amyloid-negative simulation data using a model pretrained on the simulated [$^{\text{18}}\text{F}$]FDG dataset. Markers were plotted at 10-iteration intervals from 10 to 100 for MLEM and MAPEM, at increments of 0.2 in $\lambda_\text{DPS}$ from 0.4 to 2.0 for DPS, at $\lambda_\text{DDS}$ of 0.1, 1.0, 5.0, 10.0, 25.0, 50.0, 75.0, 100.0, 150.0, and 200.0 for DDS, and at $T'$ of 20, 40, 60, 80, 100, 200, 400, 600, 800, and 1000 for the proposed method.}
\label{fig:contrast-cv}
\end{figure}

  \begin{figure*}[!t]
  \centering
  \includegraphics[width=0.9\textwidth]{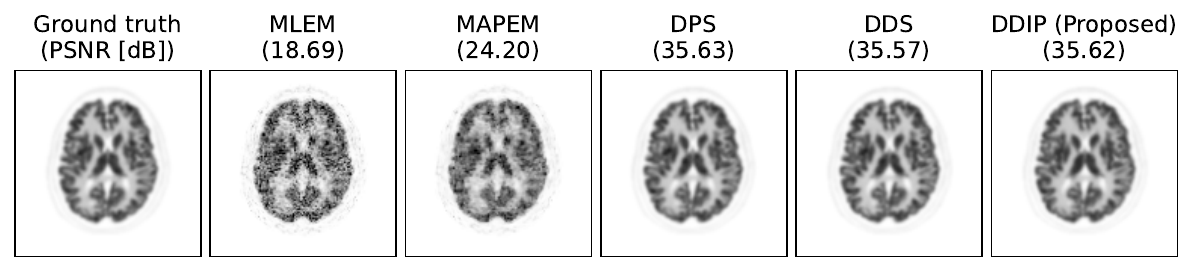}
  \caption{Simulation results of the simulated [$^{\text{18}}\text{F}$]FDG data under the ID condition.}
\label{fig:IDresults}
\end{figure*}

\subsection{Clinical brain PET data studies}
For the clinical data experiment, we evaluated the performance of the proposed method on two evaluation datasets, [$^{\text{18}}\text{F}$]FDG and [$^{\text{18}}\text{F}$]Florbetapir, both acquired on Siemens Biograph mMR scanners at different sites. We used a pretrained score function based on the early-10-mins [$^{\text{18}}\text{F}$]MK-6240 tau scans acquired from the GE Discovery MI scanner, along with the corresponding T1-weighted MR images. Note that these early-time-frame images mainly reflected cerebral blood flow information. A detailed description was provided in \cite{Gong2024}. A total of 116 PET-MR datasets from participants with normal cognition, mild cognitive impairment, or Alzheimer’s disease were employed for training. 

During testing, we used 10 datasets from the Monash DaCRA fPET-fMRI [$^{\text{18}}\text{F}$]FDG dataset \cite{Jamadar2022}, and one healthy control [$^{\text{18}}\text{F}$]Florbetapir dataset, provided by Markiewicz et al \cite{pawel}. Dynamic [$^{\text{18}}\text{F}$]FDG scans were acquired over 90 minutes following a dose of approximately 238 MBq, with the 80–90 minute frame used for testing, whereas dynamic [$^{\text{18}}\text{F}$]Florbetapir scans were acquired over 60 minutes following an injection of approximately 370 MBq \cite{Lane2017}, with the 40–60 minute frame used for testing. These clinical datasets represented prototypical OOD scenarios, as their data distributions differed from that of the pretraining dataset, which was based on [$^{\text{18}}\text{F}$]MK-6240 scans.

Low‐dose PET scans were simulated by downsampling the list-mode data to $1/10$ of the original counts. Corresponding T1‐weighted MR images were acquired on the same scanners and co-registered to the PET images using Advanced Normalization Tools in Python (ANTsPy) library \cite{tustison2021antsx}. The reconstructed images had dimensions of $344 \times 344 \times 127$ with a voxel size of $2.08 \times 2.08 \times 2.03\text{ mm}^3$. To reduce GPU memory requirements and accelerate computation, the volumes were cropped to $128 \times 128 \times 83$. Scatter and random corrections were estimated using a voxel-driven scatter model and maximum-likelihood methods, respectively. Attenuation correction was performed using MR-based $\mu$ maps. The system matrix was generated using the Siddon algorithm with point spread function modeling. All corrections and system matrix computations were implemented using the NiftyPET package \cite{Markiewicz2018}.

Regarding quantification, for the [$^{\text{18}}\text{F}$]FDG dataset, contrast recovery in the putamen region was adopted and calculated as
 \begin{equation}
\label{eq:cr}
\text{CR}=\frac{\text{ROI}_{t}}{\text{ROI}_{full}},
\end{equation}
where $\text{ROI}_{t}$ and $\text{ROI}_{full}$ denote the putamen regions of interest (ROIs) in target and full-dose images, respectively. The standard deviation (std) values were calculated in the white matter regions. ROIs for the putamen and white matter were delineated by a radiological technologist for each of the 10 subjects. For [$^{\text{18}}\text{F}$]Florbetapir data, the \%contrast and CV were computed from MR‑derived ROIs, with the full‑dose image used as the ground truth.

 \subsection{Comparison algorithms}
We compared the proposed DDIP-based reconstruction method with MLEM, MAPEM using the relative difference penalty \cite{Nuyts2002}, and DPS- and DDS-based reconstruction methods \cite{Chung2023, Chung2024, Singh2024}. Although DDIM sampling with pretrained score functions is not a reconstruction-based method, we included it as a reference to illustrate the outputs generated by the pretrained network. These results can help illustrate the distribution learned by the pretrained model.

For the MAPEM method, the relative difference penalty is
\begin{equation}
\label{eq:rdp}
R(\bm{x}) = \sum_{j}\sum_{k\in{N_j}}\frac{(x_j-x_k)^2}{(x_j+x_k)+\gamma\left|x_j-x_k\right|},
\end{equation}
where $\gamma$ is a hyperparameter that controls the shape of the function. We used the default setting for clinical PET scanners with  $\gamma=2$ \cite{Miwa2023}. For the DPS-based reconstruction, the score $ \nabla_{\bm{x}_t} \log p\bigl(\bm{y} | \bm{x}_t\bigr)$ is approximated by
\begin{equation}
\label{eq:dps}
\nabla_{\bm{x}_t}\log p\bigl(\bm{y} | \hat{\bm{x}}_0 \bigr)\approx\lambda_\text{DPS} \frac{\hat{\bm{x}}_0}{\bm{S}}\left(\bm{A}^T \frac{\bm{y}}{\bm{A}\hat{\bm{x}}_0}-\bm{S}\right)\frac{\partial\hat{\bm{x}}_0}{\partial\bm{x}_t},
\end{equation}
where $\hat{\bm{x}}_0 / \bm{S}$ is a preconditioning term and $\lambda_\text{DPS}$ is a step size.

The update of DDS-based reconstruction is
\begin{align}
\label{eq:dds}
\hat{\bm{x}}_0^{(k+1)}&=\hat{\bm{x}}_0^{(k)}\notag\\
&+\frac{\hat{\bm{x}}_0^{(k)}}{\bm{S}}\left[\bm{A}^T \frac{\bm{y}}{\bm{A}\hat{\bm{x}}_0^{(k)}}-\bm{S}-2\lambda_\text{DDS}\left(\hat{\bm{x}}_0^{(k)}-\hat{\bm{x}}_0^{(0)})\right)\right],
\end{align}
where $\hat{\bm{x}}_0^{(0)}=\hat{\bm{x}}_0$, $k=0,1,\ldots,K-1$ denotes the number of optimization steps, $\hat{\bm{x}}_0^{(k)} / \bm{S}$ is a preconditioning term and $\lambda_\text{DDS}$ is a step size. In this paper, we set the number of diffusion sampling steps to 200 and $K=5$. 

For fair comparison, the same pretrained score functions used in the proposed method were also employed for DDIM sampling and for DPS- and DDS-based reconstructions.

 \begin{figure*}[!t]
  \centering
  \includegraphics[width=\textwidth]{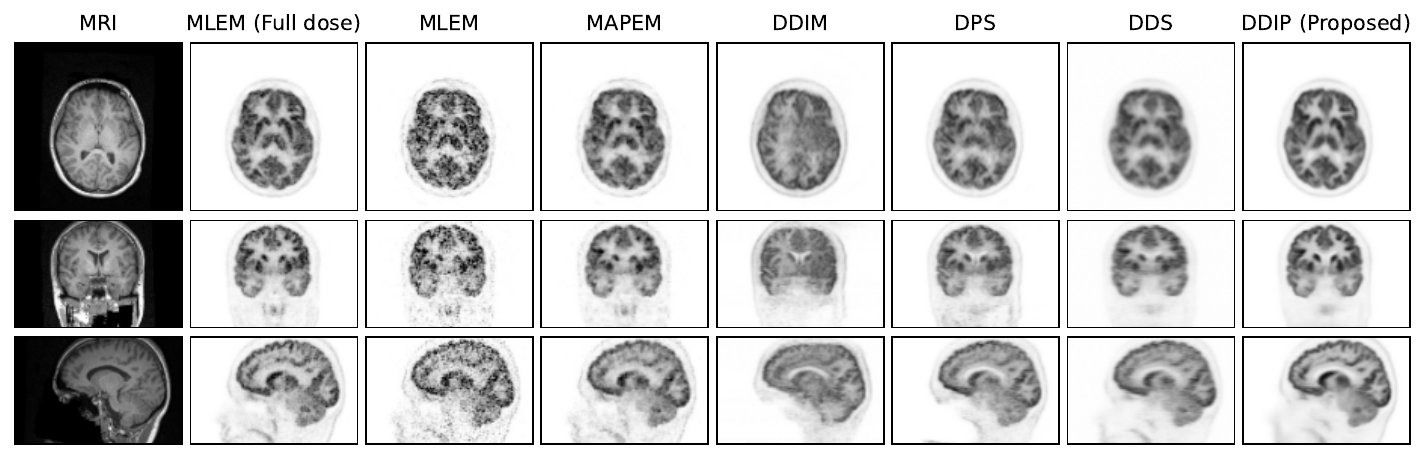}
  \caption{Reconstruction results of three orthogonal slices of the clinical [$^{\text{18}}\text{F}$]FDG data using a diffusion model pretrained on the early-10-mins [$^{\text{18}}\text{F}$]MK-6240 tau datasets.}
\label{fig:fdg_img}
\end{figure*}

 \begin{figure*}[!t]
  \centering
  \includegraphics[width=\textwidth]{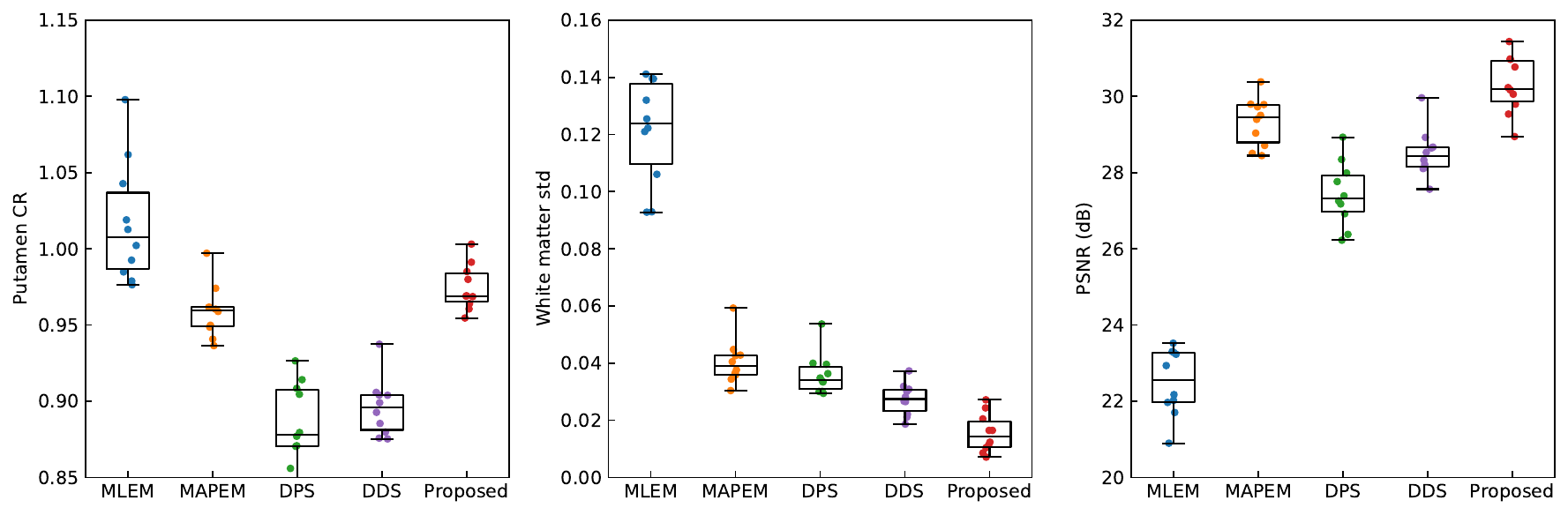}
  \caption{Box plots for putamen CR (left), white matter std (center), and PSNR (right) of clinical [$^{\text{18}}\text{F}$]FDG data reconstructed from different methods. Individual data points are overlaid on the box plots.}
\label{fig:fdg_box}
\end{figure*}

\section{Results}
\subsection{Brain phantom simulation study}
An ablation study was conducted to assess the contributions of key components, such as diffusion sampling strategy, LoRA, and anatomical prior to the proposed DDIP-based reconstruction. The evaluated component combinations are listed in Table \ref{table_ablation}. For a fair comparison, the cDIP-based reconstruction baseline was implemented using the same HQS algorithm as the proposed method. As shown in Fig. \ref{fig:ablation}, the proposed method, which employs the diffusion sampling strategy, outperforms cDIP. The ablation study also highlights the importance of anatomical information and LoRA adaptation in DDIP-based reconstruction, as their inclusion improves image quality.
Figs. \ref{fig:lora-img} illustrate the reconstructed images and corresponding effects on PSNR measures for various values of the LoRA parameter $r$. The PSNR remained stable for $r$ from 4 to 12, but gradually declined when $r > 12$. Based on these observations, we set $r=4$, which achieved the best performance with the fewest trainable parameters.

Fig. \ref{fig:simu_beta_img} represents the reconstructed results for different starting times $T'$. The proposed method successfully generated the reconstructed images for all $T'$ values; however, spatial errors increased when $T'$ was either small or large. Fig.\ref{fig:beta} presents the impact of the hyperparameter $\beta$ and starting time $T'$ on reconstruction quality, measured by PSNR averaged over 10 independent realizations. The highest PSNR was achieved at  $\beta=0.01$ and $T'=200$, with PSNR remaining relatively stable for $\beta$ values between $0.001$ and $0.1$. Based on this experiment, we used $\beta=0.01$ and $T'=200$ for the subsequent clinical-data evaluation.

Fig. \ref{fig:simu_img} presents the reconstructed results for different methods. Result of the DDIM method demonstrates that the pretrained score function effectively captured the distribution of the simulated [\textsuperscript{18}F]FDG contrast. Traditional MLEM and MAPEM suffered from high image noise under low-count conditions. In contrast, DPS, DDS and the proposed method produced images with lower image noise; the proposed method more accurately preserved gray matter structures and better recovered gray–white matter contrast compared to DPS and DDS. 

To assess the uncertainty and variability of the proposed method, voxel-wise mean, bias, and standard-deviation images were computed over 10 i.i.d. realizations, as shown in Fig. \ref{fig:simu_mean_bias_std_img}. The bias maps were computed as the voxel-wise difference between the mean images and the ground truth. Compared to MLEM and MAPEM, DPS, DDS and the proposed method exhibited lower spatial standard deviations. However, DPS and DDS showed higher spatial bias. In contrast, the proposed method offers a more favorable trade-off between variance and bias. Fig.~\ref{fig:contrast-cv} shows the mean \%contrast-CV tradeoff curves. The proposed method achieved a better tradeoff curve than the other methods. These quantitative results demonstrate that the proposed DDIP-based reconstruction outperformed both traditional iterative and state-of-the-art diffusion model-based reconstruction methods.

We conducted an ablation study under an ID scenario using a [$^{\text{18}}\text{F}$]FDG phantom for testing. Fig.~\ref{fig:IDresults} shows the reconstructed results for different methods. The proposed method performed comparably to other diffusion-based reconstruction methods, such as DPS and DDS. The results indicate that the proposed method primarily improves performance under OOD conditions rather than providing additional improvements in the ID setting.

\subsection{Clinical brain PET data studies}
Figure~\ref{fig:fdg_img} shows reconstructed images of the clinical [\textsuperscript{18}F]FDG data using different methods. The DDIM approach produced an early 10-minute [$^{\text{18}}\text{F}$]MK-6240 tau PET contrast that closely aligned with the structural features of the MR prior. DPS, DDS and the proposed method yielded images with lower noise levels compared to MLEM and MAPEM. The reasonable performances of DPS and DDS may be due to the relatively close distributional similarity between the early 10-minute [$^{\text{18}}\text{F}$]MK-6240 dataset used for training and the [\textsuperscript{18}F]FDG clinical data, as suggested by the visual resemblance between the DDIM outputs and the FDG images.

 \begin{figure*}[!t]
  \centering
  \includegraphics[width=\textwidth]{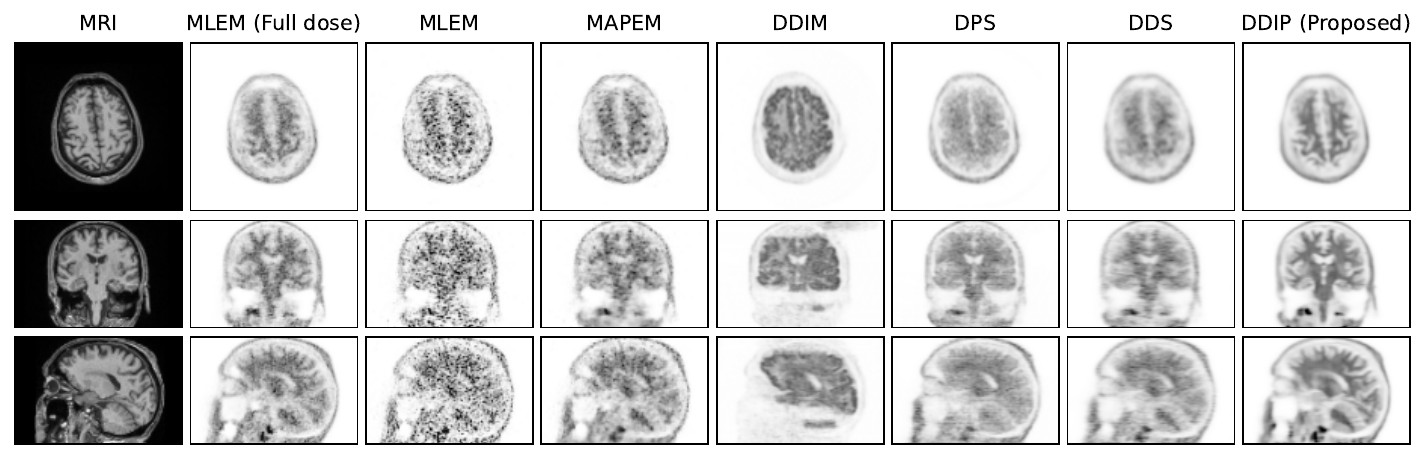}
  \caption{Reconstruction results of three orthogonal slices of the clinical[$^{\text{18}}\text{F}$]Florbetapir data using a diffusion model pretrained on the early-10-mins [$^{\text{18}}\text{F}$]MK-6240 tau datasets.}
\label{fig:amyloid_img}
\end{figure*}

Fig. \ref{fig:fdg_box} shows the putamen CR, white matter std, and PSNR across 10 individual subjects. The mean putamen CR values for MLEM, MAPEM, DPS, DDS, and the proposed method were 1.017, 0.959, 0.872, 0.896, and 0.975, respectively; the corresponding mean white matter std values were 0.121, 0.040, 0.036, 0.027, and 0.015, respectively; and the corresponding mean PSNRs were 22.50, 29.33, 27.44, 28.51, and 30.29, respectively. The proposed method achieved the mean putamen CR close to 1.0, the lowest white matter std, and the highest PSNR, demonstrating its ability to preserve quantitative accuracy while effectively suppressing statistical noise.

Fig. \ref{fig:amyloid_img} shows the reconstructed results on the clinical [$^{\text{18}}\text{F}$]Florbetapir data for different methods. The results were consistent with the simulation and clinical [\textsuperscript{18}F]FDG results. In addition, DPS and DDS exhibited streak artifacts along the axial direction, breaking continuity between slices. In contrast, the proposed method produced a spatially consistent image throughout the volume. Table \ref{table_amyloid} shows the results of \%contrast and CV for different methods. The proposed method outperformed the other comparison methods in terms of contrast and image noise.

The computational (reconstruction) times on the clinical brain PET datasets were 74.8, 51.5, and 114.4 minutes for DPS, DDS, and the proposed method, respectively.

\section{Discussion}
In this study, we proposed a diffusion model‐based, anatomical prior‐guided PET reconstruction method inspired by the DDIP framework. Both PET reconstruction and diffusion model optimization are computationally intensive, requiring substantial time and memory. Jointly optimizing the network with forward projection operations poses significant memory challenges, particularly for large-scale networks with fully 3D projections. To overcome this, we adopted the HQS algorithm to decouple network optimization from PET reconstruction, enabling a more memory-efficient and scalable implementation. Furthermore, to accelerate reconstruction, we introduced the starting time $T'$ at which the reverse-diffusion process was initiated from an intermediate timestep.

As shown in Figs. \ref{fig:lora-img}, fine-tuning the pretrained score function without LoRA resulted in severe overfitting to the measured data and thus generated noisy images. In contrast, using LoRA suppressed catastrophic forgetting by confining updates to low-rank adaptation matrices, thereby enabling the model to generate high-quality, low-noise images. However, using a larger LoRA rank ($r > 12$) may induce overfitting. Our future work will focus on investigating whether the optimal LoRA rank $r$ varies with different imaging protocols.

\begin{table}[!t]
\centering
\caption{\%contrast, CV, and PSNR of the Clinical [$^{\text{18}}\text{F}$]Florbetapir Data for Different Methods.}
\label{table_amyloid}
\setlength{\tabcolsep}{3pt}
\begin{tabular}{lccc}
\toprule
Methods& 
\%contrast& 
CV&
PSNR (dB)\\\midrule
MLEM& 
94.82\%& 
0.736& 
20.50\\
MAPEM& 
90.54\%& 
0.300&
26.87\\
DPS& 
51.08\%&
0.230&
25.29\\
DDS& 
75.22\%& 
0.228&
27.76\\
DDIP (Proposed)& 
99.97\%& 
0.220&
29.07\\
\bottomrule 
\end{tabular}
\end{table}

In the context of PET image denoising, diffusion models have demonstrated superior generalizability compared to U-Net and generative adversarial networks, exhibiting robustness across varying noise levels and scanner types \cite{Gong2024, Yu2025}. Although DPS and DDS achieved superior reconstruction performance under the ID condition, their generalization remains limited in more challenging OOD scenarios, such as contrast inversion, as shown in the simulation study. This limitation arises because DPS and DDS rely on frozen pre-trained score models. When the training data do not encompass the tracer distribution present at test time, the score function becomes mismatched with the test data, which limits adaptation to OOD conditions. In contrast, the proposed DDIP-based reconstruction method produced higher-quality images than DPS and DDS. This improvement stems not only from the strong generative capacity of diffusion models but also from the fine-tuning step described in Eq.~(\ref{eq:hqs2}), which adapts the score function to the test data rather than relying on a frozen pre-trained score as in DPS and DDS. By aligning the pretrained score function with the measured PET data, this step effectively shifts the model's prior toward the target (test) domain, enhancing reconstruction quality under severe OOD conditions.

The voxel-wise bias maps in Fig.~\ref{fig:simu_mean_bias_std_img} illustrate reconstruction errors relative to the ground truth. DPS and DDS substantially underestimated tracer uptake across the brain, whereas the proposed method produced unbiased contrast closely matching the ground truth. In addition, the low variability observed in the standard deviation maps highlights the stability of the proposed approach. Clinical studies using two test datasets acquired on tracers and scanners different from those used during training further demonstrate the method’s robustness under OOD conditions. In the [$^{\text{18}}\text{F}$]FDG experiment, whose distribution closely resembled the early 10-minute [$^{\text{18}}\text{F}$]MK-6240 scans used for pertaining, DPS and DDS showed reasonable performance but introduced hallucinated structures. In contrast, the proposed method produced more anatomically accurate reconstructions. In the [$^{\text{18}}\text{F}$]Florbetapir experiment, it achieved the highest \%contrast and the lowest noise. Together, these simulation and clinical results suggest that the proposed DDIP-based reconstruction framework generalizes effectively across a wide range of tracers and scanner types.

One limitation of this study is that the ground truth in the clinical evaluations was based on full-dose images, which still contain statistical noise. As a result, some putamen CR values slightly exceeded 100\%, but it remains unclear whether this reflects true overestimation or inherent noise in the reference images. Another limitation is the use of 2D network architectures. Notably, despite this constraint, the proposed method exhibited minimal slice-to-slice variability, with no perceptible discontinuities upon visual inspection. Future work will explore memory- and computation-efficient 3D network architectures to further enhance performance. Given that the [$^{\text{18}}\text{F}$]Florbetapir evaluation is limited to a single subject, future work will include a larger, multi-subject cohort. We will also extend the clinical evaluation to whole-body PET datasets to assess the method’s generalizability across diverse anatomical regions and radiotracers.

\section{Conclusion}
In this work, we proposed a DDIP-based and anatomical prior-guided PET image reconstruction framework. Evaluations based on simulation and clinical datasets showed that the proposed method produced more accurate PET images than other reference reconstruction methods. These results demonstrate that the proposed PET reconstruction can generalize robustly across different tracer distributions and scanner configurations, offering a versatile and efficient reconstruction framework for low‑dose PET imaging.


\bibliographystyle{IEEEtran}
\bibliography{DDIP_ref}

\end{document}